\documentclass[epj,final]{svjour}
\usepackage{graphicx}
\newcommand{\dd}{\mathrm d}
\usepackage{amsmath}
\usepackage{latexsym}
\usepackage[dvips]{epsfig}
\newcommand{\beq}{\begin{equation}}
\newcommand{\eeq}{\end{equation}}

\newcommand{\un}[1]{\ensuremath{\unskip\,\mathrm{#1}}}

\begin{document}

\title{Structure and interaction potentials in solid-supported
lipid membranes studied by X-ray reflectivity at varied osmotic pressure}

\titlerunning{X-ray reflectivity of lipid lamellar stacks under osmotic pressure}

\author{Ulrike Mennicke\thanks{\email{ulrike.mennicke@gmx.net}} \and
Doru Constantin\thanks{\email{constantin@lps.u-psud.fr} Permanent
address: Laboratoire de Physique des Solides, Universit\'{e}
Paris-Sud, B\^{a}t. 510, 91405 Orsay Cedex, France.} \and Tim
Salditt\thanks{\email{tsaldit@gwdg.de}}}

\institute{Institut f\"{u}r R\"{o}ntgenphysik, Friedrich-Hund-Platz
1, 37077 G\"{o}ttingen, Germany}


\date{\today}

\abstract{Highly oriented solid-supported lipid membranes in
stacks of controlled number $N \simeq 16$ (oligo-membranes) have
been prepared by spin-coating using the uncharged lipid model
system 1,2-dimyristoyl-sn-glycero-3-phosphocholine (DMPC). The
samples have been immersed in aqueous polymer solutions for
control of osmotic pressure and have been studied by X-ray
reflectivity. The bilayer structure and fluctuations have been
determined by modelling the data over the full $q$-range. Thermal
fluctuations are described using the continuous smectic
hamiltonian with the appropriate boundary conditions at the
substrate and at the free surface of the stack. The resulting
fluctuation amplitudes and the pressure-distance relation are
discussed in view of the inter-bilayer potential.
} 

\maketitle

\section{\label{Intro}Introduction}

A quantitative understanding of the structure, fluctuations,
interaction potential and elasticity properties of lipid
membranes, which represent model systems for biological membranes,
has been the goal of many theoretical and experimental studies.
Theoretically, they have been studied as paradigmatic examples of
quasi two-dimensional macromolecular structures governed by
bending rigidity \cite{Lipowsky}. In aqueous solution, lipid
bilayers assemble into stacks governed by distinct inter-bilayer
interactions. A number of seminal studies using high resolution
synchrotron X-rays have been published on these systems
\cite{Safinya,Petrache,petrache98,PabstPRE}. In these studies,
isotropic aqueous dispersions of multilamellar vesicles have been
studied as a function of temperature $T$ and/or osmotic pressure
$\Pi$. Detailed quantitative information on the interaction
potentials and the elasticity properties has thus been derived,
see {\em e.g.} \cite{Safinya,Petrache,petrache98}. However,
information is lost in these small-angle scattering studies due to
crystallographic powder averaging. In the analysis, assumptions
must therefore be made on the nature of the correlation functions
in the framework of linear smectic elasticity theory, leading to
the Caill\'e model \cite{Caille} and related theories, see for
instance \cite{Zhang}. In order to overcome the limitations of
powder averaging, it is desirable to work with aligned systems of
lipid bilayers \cite{Smith,Lyatskaya,VogelPRL,SaldittPRL,Liu04}.
Under the same conditions of temperature and hydration, thermal
fluctuations are not as strong for aligned systems as in bulk
studies due to the boundary condition at the flat substrate,
enabling a higher resolution in $\rho(z)$. It is also advantageous
to fit the data continuously over a large range of momentum
transfer $q$, as has been shown for isotropic solutions
\cite{PabstPRE} and for oriented stacks \cite{Liu04}, and not only
in the vicinity of the Bragg peaks arising from the multilamellar
structure. However, for aligned systems of multilamellar
membranes, satisfactory fits of the entire reflectivity curves and
the formulation of a proper statistical model as well as of a
scattering theory are still quite difficult. Note that the best
published fits are for systems consisting of monolayers at the
air-water interface, for single bilayers or for a free-floating
bilayer system developed by Fragneto and coworkers
\cite{Fragneto}. In multilamellar systems on the other hand, the
reflectivity signal is typically much more complex and structured.
As we show here, structural parameters of the bilayer, interaction
and fluctuation parameters can be deduced from these curves, and
are compared with the literature.

Most previous studies on aligned multilamellar membranes suffered from a lack of control concerning the
number of bilayers $N$, the sample homogeneity and also the bilayer hydration. Building on recent progress
in the preparation of bilayers on solid support by spin-coating \cite{spin}, we use so-called
oligo-membranes with a reduced number of bilayers $N\simeq 10-20$ resulting in very structured and well
resolved reflectivity signals. We have developed a model for the thermal fluctuations calculated for the
proper boundary conditions (rigid substrate and free upper surface), which gives the fluctuation amplitudes
$\left \langle u_n^2 \right \rangle$ of the bilayers as a function of their position in the stack,
$n=\overline{1,N}$. The values for $\left \langle u_n^2 \right \rangle$ are then inserted in the
multilamellar structure factor, along with a decreasing coverage function (see below). The density profile
$\rho(z)$ is parameterized in terms of its Fourier coefficients \cite{Li}. This approach gives for the
first time an agreement with the measured reflectivity curve of multilamellar membranes over the full range
of $q_z$ up to typically $q_z \simeq 0.7 \, \mathrm{\AA}^{-1}$ and over about seven orders of magnitude in
the measured signal.

The main experimental parameter in this study is the osmotic pressure. The classical osmotic stress (OS)
technique as developed by Parsegian and coworkers \cite{Parsegian} is widely used for the measurement of
force-distance curves in colloidal systems. The osmotic pressure imposed to a lamellar phase controls the
interaction force experienced by the membranes across the water layer by setting the chemical potential of
the water molecules in the inter-bilayer solution. Pressure-distance relations can be easily determined,
{\em e.g.} if the lamellar periodicity $d$ is measured by X-ray scattering at different pressures $\Pi$.

\begin{figure}[htbp]
\begin{center}
\includegraphics[width=8.5cm]{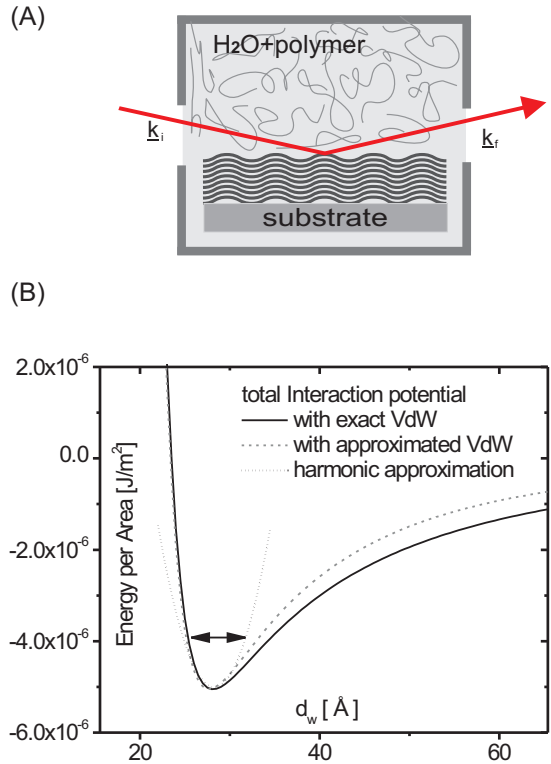}
\caption{ A): Sketch of the experimental setup where a
solid-supported membrane stack is hydrated in an aqueous polymer
solution. Arrows indicate the incoming and scattered X-ray beams.
B): Interaction potential with van der Waals contribution
according to Fenzl \cite{Fenzl} (solid line) and with general
approximation (dashed). The (dotted) parabola illustrates the
harmonic approximation to the potential which enters in the
smectic elasticity theory. The arrow shows a typical fluctuation
amplitude of a membrane inside a 16 bilayer stack. \label{sketch}}
\end{center}
\end{figure}

In this study we use a variant of the OS technique where the
oriented bilayers are put in direct contact with the osmotic
stress solution \cite{Brotons03}. We have verified that the high
molecular weight polymer with a radius of gyration larger than $d$
does not penetrate the lamellar phase. We emphasize that the
osmotic pressure is one of the most important parameters in
biomolecular systems, since biomolecular assemblies in the cell
are mostly exposed to varying $\Pi$ while $T$ is often constant.
Therefore, it is of great importance to study bilayer structure
and elastic properties such as bilayer bending rigidity $\kappa$
or bilayer-bilayer interaction parameters as a function of osmotic
pressure $\Pi$.

From the relation between the osmotic pressure $\Pi$ and the
lamellar spacing $d$ the inter-bilayer interaction potentials can
be determined. It is generally accepted that in charge neutral
systems two main molecular interaction forces are dominant, in
addition to the effective attractive interaction by osmotic
pressure~: a repulsive hydration potential $f_{\mathrm{hyd}}
(d_w)$ and the attractive van der Waals potential
$f_{\mathrm{vdW}} (d_w)$ so that the total interaction in
$\mathrm{J/m}^2$ is given by $f(d_w)=f_{\mathrm{hyd}}
+f_{\mathrm{vdW}} +\Pi d_w$, defining the equilibrium distance
(water layer thickness) $d_w$ as illustrated in Fig. \ref{sketch}
B). As discussed below, it is important to take the correct form
for $f_{\mathrm{vdW}}$ as derived in \cite{Fenzl}, without the
conventional half-space approximation. It can be argued that
steric (Helfrich) repulsion forces have to be added to the
molecular forces in a mean field approach. Here, however, we will
assume that thermal fluctuations in thin films of relatively stiff
phospholipids do not have a significant effect on the
inter-bilayer interactions, in particular since the flat boundary
suppresses long range fluctuations in the film. The paper is
organized as follows~: After this introduction, section
\ref{Exper} presents some experimental details while the
statistical model and data analysis are presented in section
\ref{Model}. Section \ref{Reflect} presents the results, followed
by a section on the interaction potentials and the conclusions in
section \ref{Conc}.

\section{\label{Exper}Materials and Methods}

\subsection{Samples and Environment}

Highly oriented oligo-membranes were prepared using the
spin-coating method \cite{spin}. The uncharged lipid
1,2-di\-my\-ri\-stoyl-sn-glycero-3-phosphocholine (DMPC) was bought from
Avanti (Alabaster, AL, USA) and used without further purification.
The lipid was dissolved in chloroform at a concentration of 10
mg/ml. An amount of 100 $\mu \mathrm{l}$ of the solution was
pipetted onto carefully cleaned silicon substrates of a size of 15
$\times$ 25 mm$^2$ cut from standard commercial silicon wafers.
The substrate was then immediately accelerated to rotation (3000
rpm), using a spin-coater. After 30 seconds the samples were dry
and subsequently exposed to high vacuum to remove any remaining
traces of solvent. The samples were then stored at $4 \,
^{\circ}\mathrm{C}$ until the measurement. For the X-ray
measurements the samples were hydrated in a stainless steel
chamber \cite{VogelPRL} with kapton windows, which can be filled
with water or with polymer solution to control the level of
hydration by osmotic pressure, see the sketch in Fig.
\ref{sketch}A. Temperature was controlled by an additional outer
chamber at $T=40 \,^{\circ}\mathrm{C}$. The polymer
polyethyleneglycol (PEG) of molar weight 20000 Da was bought from
Fluka and used without further purification. PEG was dissolved in
ultrapure water (Millipore, Billerica, Mass.) at the
concentrations 1.5 \%, 2.9 \%, 3.6 \%, 5.8 \%, 9 \%, 12.1 \%, 14.2
\% and 25 \% (wt. percent). The corresponding osmotic pressure
values were taken from the literature. The data was obtained from
the web site of the Membrane Biophysics Laboratory at the Brock
University in Canada (http://aqueous.labs.brocku.ca/osfile.html).
The value for the osmotic pressure of PEG 20000 solutions is only
available at $20\,^{\circ}\mathrm{C}$. At $40
\,^{\circ}\mathrm{C}$, temperature at which the experiments were
performed, the pressure can be expected to be somewhat lower.
However, the temperature coefficient is small
\cite{Parsegian,StanleyStrey}, and the corresponding discrepancy
smaller than the error bars in Fig \ref{fig:piofd} below, which is
used in the analysis of the interaction forces.

An important issue in our method of direct contact of the lamellar
phase with the polymer solution is that of possible
interpenetration of the multilamellar phase. Experiments on
polymer containing lyotropic lamellar phases
\cite{Ligoure97,Ligoure99} have shown that polymers can enter the
water layer in between charged bilayers even if the radius of
gyration is larger than the water layer. However, in the present
case of neutral polymer in neutral bilayers, the amount of polymer
in between the lamellae is negligible. The experimental proof is
given by (a) the density profile, which shows no deviation from
the water density in between the bilayers, and (b) the force
distance curve itself which shows no indication of such an effect.

\subsection{X-ray experiment}

\begin{figure}[htbp]
\includegraphics[width=8.5cm]{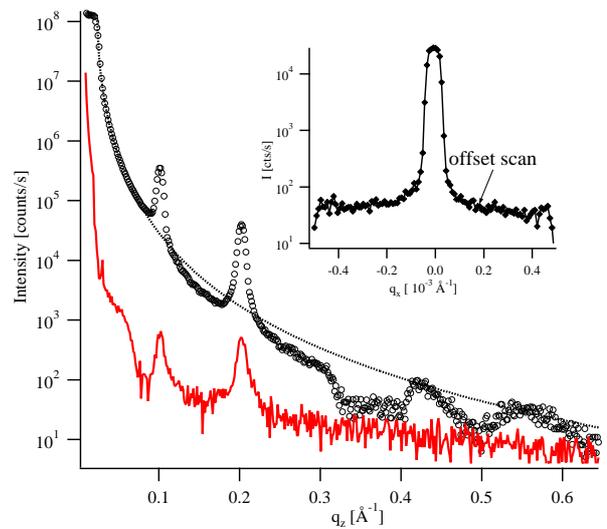}
\caption{Specular reflectivity scan (open symbols) and offset scan (red solid curve) for a sample immersed
in pure water. The "true specular" contribution (see Appendix) is given by the difference of the two
curves. In dotted line we also show the  Fresnel reflectivity profile corresponding to the same critical
angle $q_c$. Inset~: Rocking scan at the position of the second Bragg peak. The arrow shows the $q_x$
position of the offset scan. \label{fig:refl_pure_water}}
\end{figure}

The X-ray reflectivity measurements presented here were carried out at the bending magnet beamline D4 of
HASYLAB/DESY in Hamburg, Germany. At D4, a single-reflection Si(111) monochromator was used to select a
photon energy of 19.92 keV, after passing a Rh mirror to reduce higher harmonics. The chamber was mounted
on the $z$-axis diffractometer, and the reflected beam was measured by a fast scintillation counter
(Cyberstar, Oxford), using computer-controlled aluminum absorbers which attenuate the beam at small $q_z$
to prevent detector saturation. Incident and exit beams were defined by a system of several motorized
slits. The data is corrected for decreasing electron ring current and the diffuse contribution by
subtraction of an offset scan. Finally, an illumination correction is performed. A typical measurement
(reflectivity and offset scans) is shown in Figure \ref{fig:refl_pure_water}, along with the corresponding
Fresnel reflectivity. The inset shows a rocking scan on the second Bragg peak, illustrating the separation
between "true specular" and diffuse components.

\section{\label{Model}Model and data analysis}

\subsection{Reflectivity}

In the semi-kinematic approximation the reflectivity of a structured interface can be expressed by the
so-called master-formula of reflectivity \cite{AlsNiels} as~:
\begin{equation}\label{master}
R(q_z)=R_{F}(q_z) \cdot \left | \frac{1}{\rho_{12}}\int_{-\infty}^{\infty} \frac{\dd \rho(z)}{\dd z}
e^{iq_zz} \dd z \right |^2 \, ,
\end{equation}
where $R_{F}$ denotes the Fresnel reflectivity of the sharp interface and $\rho(z)$ is the intrinsic
electron density profile, whereas $\rho_{12}$ is the total step in electron density between the two
adjoining media. The electron density profile of a solid-supported oligo-membrane stack, consisting of $N$
membranes in water, can then be written as~:

\begin{equation}
\label{rhoofz} \rho(z)=\frac{\rho_{12}}{2} \, \mathrm{erfc}\left ( \frac{z+d_0}{\sigma} \right ) +
\sum_{n=0}^{N-1}\rho_0(z-nd+u_n) \, ,
\end{equation}
with $\mathrm{erfc}(z)$ being the complementary error function and $\sigma$ the rms substrate roughness.
$\rho_{12}=\rho_{\mathrm{Si}} - \rho_{\mathrm{H}_2\mathrm{O}}$ is the difference in electron density
between the substrate and water, $d_0$ is the distance between the substrate and the midpoint of the first
bilayer and $\rho_0(z)$ is the electron density profile of one bilayer in the stack. Thermal membrane
fluctuations are considered in terms of the displacement function $u_n=u({\mathbf r}_{||},z=nd)$ of the
position of the $n$-th membrane from its average position $z=nd$ along $z$. Replacing the electron density
profile (\ref{rhoofz}) into (\ref{master}) and taking the ensemble average yields~:
\label{R}
\begin{align*} \label{R}
& R(q_z)  =  R_F(q_z)\cdot \Bigg [ \, \mathrm{e}^{-q_z^2\sigma^2}  -    \\
& -2i \cdot \mathrm{e}^{-\frac{q_z^2\sigma^2}{2}}
\sum_{n=0}^{N-1}f(n)\left( Ff(q_z) \cdot \sin(q_z(d_0+nd))\,
\mathrm{e}^{-\frac{q_z^2}{2}\langle u_n^2\rangle} \right)\\
\nonumber &  +|Ff(q_z)|^2 \cdot  \sum_{m,n=0}^{N-1} f(n) f(m) \,
\mathrm{e} ^{-iq_zd(m-n)} \, \mathrm{e}^{-\frac{q_z^2}{2} (\langle
u_m^2\rangle + \langle u_n^2\rangle)} \Bigg ] . \nonumber
\end{align*}
The first summand represents the reflectivity of the substrate. The second is a cross-term and represents
interference effects between the substrate and the membrane stack. The third summand is the product of the
form factor $\left | Ff(q_z) \right | ^2$ containing the structural information about one bilayer in the
stack, and the structure factor, representing the periodic structure in the stack. The fluctuations are
described by the correlation function $\langle u_n u_m \rangle $. In specular reflectivity only the
self-correlation function $\langle u_n^2\rangle$ is important (see Appendix).

\subsection{Correlation Function}

The self-correlation function of the membrane fluctuations can be calculated from linear smectic elasticity
theory based on a continuous model \cite{poniewierski}. The complete theory and calculations are described
in \cite{constantin}. Here only the essentials, which are important for specular reflectivity shall be
given. The linearized free energy can be written as a function of the displacement $u({\mathbf r}_{||},z)$ as~:
\begin{equation}
F = \frac{1}{2} \int_V d{\mathbf r} \left [ B \left (
\frac{\partial u({\mathbf r}_{||},z)}{\partial z} \right ) ^2 + K
( \Delta _{||} u({\mathbf r}_{||},z) ) ^2 \right ] \, ,
\label{caille}
\end{equation}
with $K=\kappa/d$ the bending modulus and B the compression modulus in the stack. We neglect the surface
tension between the lipid stack and the solvent. The discrete structure of the stack consisting of $N$
bilayers is taken into account by expanding $u$ over $N$ independent modes. Also, we are only interested in
the fluctuation amplitude at the position of the bilayer midpoints, denoted by $u_n = u(z=nd)$ for the
$n$-th bilayer. From the equipartition theorem one can calculate the correlation function of the membrane
fluctuations $\langle u_n^2 \rangle$, which reads~:
\begin{eqnarray} \label{eq:Cnn}
\langle u_n^2 \rangle = \eta \bigg( \frac{d}{\pi} \bigg)^2
 \sum_{j=1}^{N} \frac{1}{2j-1}
 \sin^2 \bigg( \frac{2j-1}{2} \pi
\frac{n}{N} \bigg)
\end{eqnarray}
with the conventional Caill\'{e} factor $\eta = \frac{\pi}{2 d^2}
\frac{k_B T}{\sqrt{B K}}$. Figure \ref{sigma}A shows the function
for a 16 membrane stack with a typical $\eta$ value for DMPC
membranes at partial hydration. In contrast to oligo-membranes
which are partially hydrated from water vapor, oligo-membranes
immersed in excess water or polymer solution exhibit defects which
result in decreasing layer coverage with increasing distance from
the substrate. This effect is evidenced experimentally by the
suppression of thickness oscillations (Kiessig fringes) in the
reflectivity curves. Thickness oscillations are typically observed
in vapor-hydrated samples but are significantly reduced or
suppressed in samples immersed in aqueous solution
\cite{spin,dewetting}. In the model we take this effect into
account by multiplying the contribution of each membrane in the
structure-factor with an empirical coverage-factor
\mbox{$f(n)=[1-(n/N)^{\alpha}]^2$}, where $\alpha$ parameterizes
the decaying density due to decreasing coverage. A typical
experimental value is $\alpha \approx 1.7$. Note that both
$\alpha$ and the number of bilayers $N$ are fit parameters.

Since the form factor $\left | Ff(q_z) \right | ^2$ of the
membranes consists of the squared Fourier transform of the
$z$-derivative of the electron density profile of the membrane, it
is convenient to express the profile in terms of normalized
Fourier coefficients \cite{Li}
\[ \rho(z)=\sum_{m=1}^{N_0}f_m
\cdot \cos \Big(\frac{2 \pi m z}{d} \Big) \cdot
\rho_{12}+\bar{\rho} - \rho_w \, , \]
\noindent with $\bar{\rho}$ being the
average electron density of the membrane stack and $\rho_w$ the
electron density of water. For DMPC, $\bar{\rho}=0.3397
\,\mathrm{e}^{-}/\mathrm{\AA}^{3}$ \cite{petrache98}, very close
to $\rho_w=0.332 \, \mathrm{e}^{-}/\mathrm{\AA}^{3}$, so that
$\left | Ff(0) \right | \simeq 0$.

\section{\label{Reflect}Reflectivity results}

\begin{figure*}[htbp]
\includegraphics[width=17.5cm]{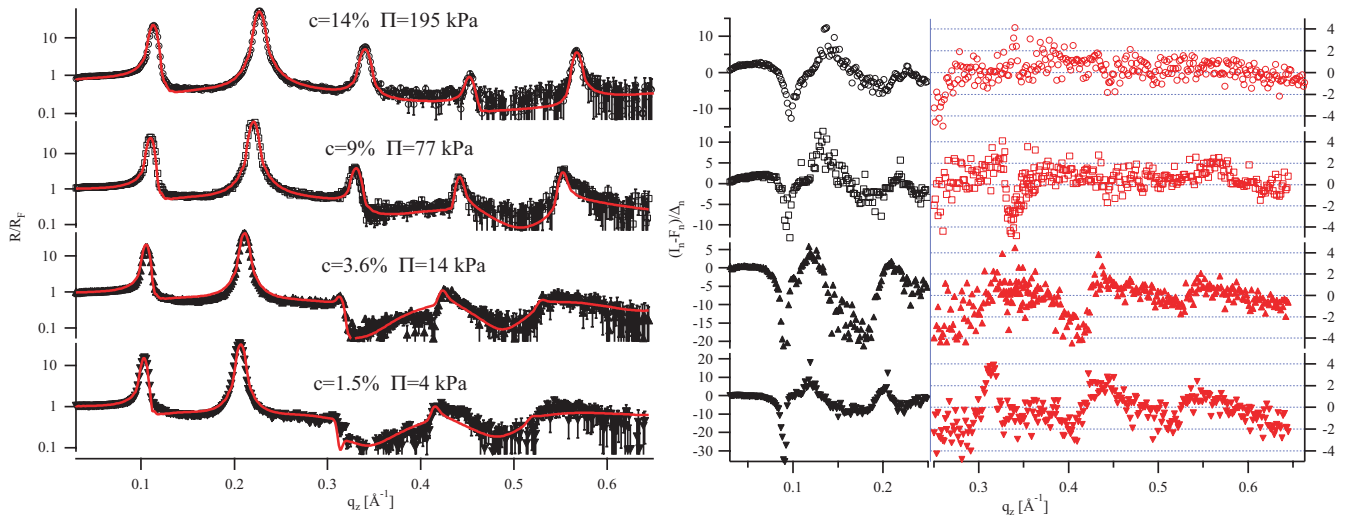}
\caption{ Left~: Symbols~: measured reflectivity curves of 16 DMPC membranes on silicon substrates under
different osmotic pressures, normalized by the corresponding Fresnel reflectivity. Solid lines~: show full
$q$-range simulations using the described model, with the density profiles given in Figure \ref{fig:dens}.
The polymer concentration and the osmotic pressure are specified for each curve. Right~: Residues of the
reflectivity data, normalized by the estimated standard deviation (see text for details). The graph is
horizontally split into two panels, with different $y$ axes~: Below $0.25 \, \mathrm{\AA}^{-1}$, the large
discrepancies correspond to systematic errors, while for $q_z > 0.25 \, \mathrm{\AA}^{-1}$ the residues are
more randomly distributed and their scaled amplitude is below 5. \label{Refl}}
\end{figure*}

\begin{figure}[htbp]
\includegraphics[width=8.5cm]{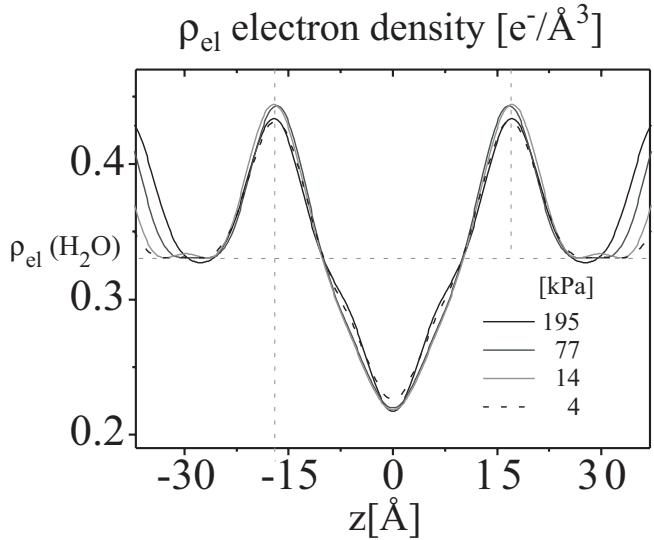}
\caption{Electronic density profile of the bilayer, as a function of the applied osmotic pressure $\Pi$.
\label{fig:dens}}
\end{figure}

Figure \ref{Refl} shows the reflectivity measurements (symbols) of
16 DMPC membranes on silicon substrates at four (out of nine
measured) different osmotic pressures. The curves have been
stacked vertically for clarity, with increasing pressure from 4
kPa (bottom) to 195 kPa (top).
The continuous lines are simulations based on the model described
above with the corresponding electron density profiles shown in
Figure \ref{fig:dens}A. The reflectivity spectra presented in
Figure \ref{Refl} were each scaled by the corresponding Fresnel
reflectivity. The error bars in the intensity at point $n$,
$\Delta I_n$, are estimated considering Poissonian statistics
(both for the raw reflectivity and for the offset scan). The error
bar in the $q_z$ direction is taken as $\Delta q_n=\Delta q=5
\times 10^{-4} \mathrm{\AA}^{-1}$, corresponding roughly to the
symbol size, and is given by the estimated precision in sample
alignment.

\begin{figure*}[htbp]
\includegraphics[width=17.5cm]{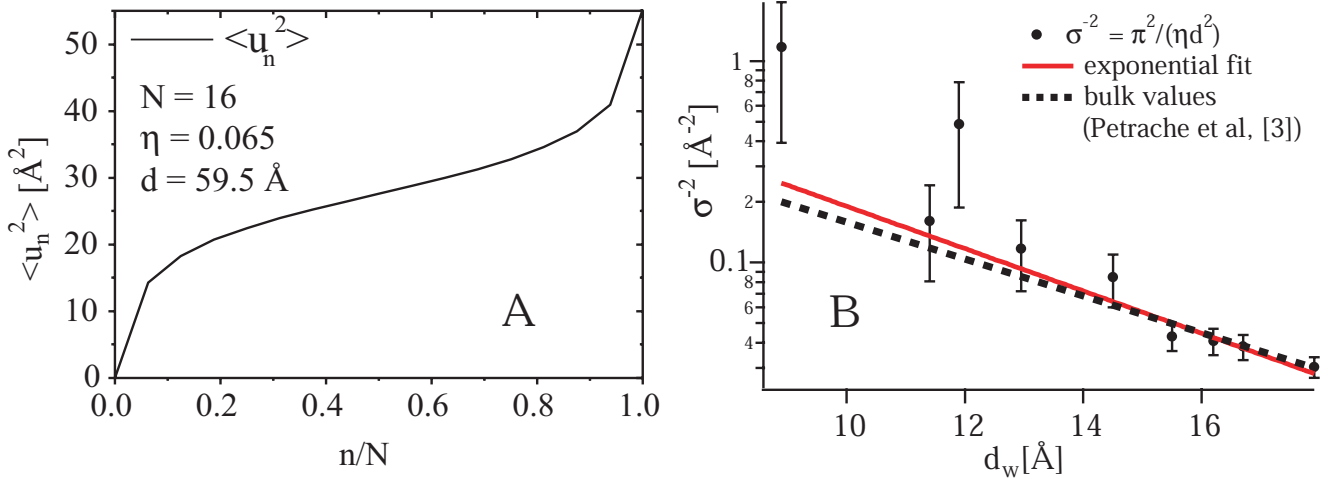}
\caption{ A) Amplitude of the bilayer position fluctuations $\left \langle u_n^2 \right \rangle$ in a
solid-supported membrane stack consisting of 16 DMPC membranes at partial hydration, controlled by an
osmotic pressure of 4 kPa (used for the model in Fig. \ref{Refl}, bottom). B) Reciprocal of the squared
fluctuation amplitudes $\sigma^{-2} (d) = \pi ^2/(\eta d^2)$, as determined from the reflectivity fits, and
exponential fit (solid line). Also shown is the exponential fit to the bulk data of Petrache {\em et al.}
\cite{petrache98}. For clarity, their experimental data points are not shown. \label{sigma}}
\end{figure*}

>From the electron density profiles one can see that the increase
of periodicity $d$ with decreasing pressure is mainly due to
changes in the water layer thickness, while the bilayer structure
is essentially invariant over the range in $\Pi$ studied, with a
headgroup spacing (distance between the two maxima in the electron
density profile) $d_{HH}=34 \pm 0.5 \, \mathrm{\AA}$, in good
agreement with the value of $34.4 \, \mathrm{\AA}$ given by
Petrache {\em et al.} \cite{petrache98}. The simulations match the
measured reflectivity curves not only at the position of the Bragg
peaks, but in the whole continuous $q$-range of the measurement.
At lower osmotic pressure $\Pi$ the higher order peaks are
suppressed due to increased thermal fluctuations, as quantified by
the above model. As an illustration, Fig. \ref{sigma}A shows the
increase in the fluctuation amplitude as a function of the
membrane index for a 16 bilayer sample.

As discussed above, the relevant parameters for the reflectivity
curves are the mean squared fluctuation amplitudes $\left \langle
u_n^2 \right \rangle$, which give access to the Caill\'{e}
parameter $\eta$. In order to compare our data to the bulk results
\cite{petrache98} we then compute the interbilayer spacing
fluctuation amplitude $\displaystyle \sigma^{2}=\left \langle
(u_n-u_{n-1})^2 \right \rangle=\eta \frac{d^2}{\pi^2}$. Soft
confinement theories predict an exponential dependence
\cite{petrache98} of parameter $\sigma^2$ with the interbilayer
distance, which can be taken as the thickness of the water layer,
given by $d_w =d-d_B$, where $d_B = 44 \, \mathrm{\AA}$ is the
thickness of the DMPC bilayer \cite{petrache98}.

Fig. \ref{sigma}B shows the reciprocal of the fluctuation
amplitudes as a function of the water spacing $\sigma^{-2} (d_w)$,
along with a fit to an exponential decay (solid line). The data
points can be fitted to an exponential function $\sigma^{-2}
\propto \exp (- d_w / \lambda_{\mathrm{fl}})$, with a decay length
$\lambda_{\mathrm{fl}}=4.1 \pm 0.9 \, \mathrm{\AA}$, comparable to
the results of the bulk study, which also reports exponential
behaviour (shown as a dashed line) with $\lambda_{\mathrm{fl}} =
5.1 \, \mathrm{\AA}$ \cite{Petrache}.

\begin{figure}[htbp]
\includegraphics[width=8.5cm]{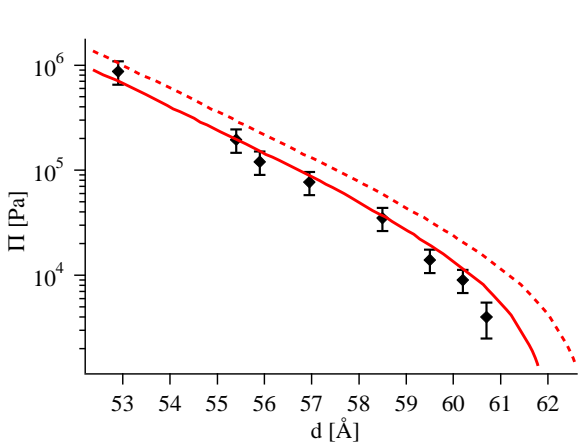}
\caption{Osmotic pressure $\Pi$ as a function of the lamellar $d$ spacing. Diamonds~: experimental data
points. Dashed line~: Fit to the bulk data of Petrache {\em et al.} \cite{petrache98}. For clarity, their
experimental data points are not shown. Solid line: same curve shifted to lower $d$ values to account for
the effect of temperature (see text for details). \label{fig:piofd}}
\end{figure}

The lamellar periodicity $d$ was measured for all values of the osmotic pressure $\Pi$, up to 870 kPa. We
show $\Pi (d)$ in Figure \ref{fig:piofd} (diamonds). For comparison, we also plotted the fit by Petrache
{\em et al.} of the bulk data, Figure 7, upper panel in their paper \cite{petrache98} (dashed line). They
performed the measurements at $30 \,^{\circ}\mathrm{C}$ and obtain $d_0=d(\Pi=0)=62.7 \, \mathrm{\AA}$,
while our experiments, performed at $40 \,^{\circ}\mathrm{C}$, yield $d_0=61.9 \, \mathrm{\AA}$. This
deviation is in agreement with recent data of $d_0(T)$ for DMPC \cite{Pabst04}. A quick test can be
performed by shifting their curve by $0.8 \, \mathrm{\AA}$ to lower $d$ values (solid line). The agreement
is good, but a more detailed analysis is obviously needed for a meaningful comparison.

\section{\label{Potentials}Interaction potentials}

We saw above that our data can be brought to agreement with the
bulk equation of state $\Pi(d)$ by the $0.8 \,
\mathrm{\AA}$ shift, which can be attributed to the temperature
difference. Consequently, the interaction potentials derived from
the bulk and the present data will also be identical or at least
similar. However, the choice of the functional forms for the
potentials, the geometric partitioning of the bilayer to calculate
the van der Waals part and the choice of the bending rigidity
$\kappa$ can all be debated. We therefore first give a brief
discussion of the different interaction potentials used for
neutral lipid membranes, and then present results based on
modelling the equation of state $\Pi(d)$.

The hydration potential is usually empirically described by an
exponential function of the water layer thickness $d_w$
\cite{Rand1}~:
\begin{equation}
f_{\mathrm{hyd}}(d_w)=H_0 \exp(-d_w /\lambda ) ~, \label{hydr}
\end{equation}
with a prefactor of the order of \mbox{$H_0=$
$k_BT/\mathrm{\AA}^2$} and a decay length on the order of a few
Angstroms \mbox{$\lambda=1-2$ \AA}.
For the van der Waals potential,  Petrache et al. \cite{Petrache},
used the following expression and  geometric convention:
\begin{equation}
V_{vdW}(d_w)= \frac{H_{vdW}}{16 \pi} \left [ \frac{1}{{d_w}^2}-
\frac{2}{(d_w+d_B)^2}+\frac{1}{(d_w+2d_B)^2} \right ] ~,
\label{vdW_petrache}
\end{equation}
where $d_w$ is the water layer and $d_B=d-d_w$ the bilayer
thickness. The expression should be regarded as an approximation
to the result of a more detailed treatment, as discussed by Fenzl
\cite{Fenzl}. Accordingly, the potential should be calculated from
\begin{eqnarray} \label{VdWexakt}
F_{\mathrm{vdW}}(d_h,T)=0.9\frac{k_B T}{8 \pi
d_h^2}\sum_{n=0}^{\infty \prime}
\int_{r_n}^{\infty}\dd x \, x \\
\ln \left [ 1-\left (
\frac{\Delta_n(1-\exp(-ax/d_h))}{1-\Delta_n^2 \exp(-ax/d_h)}\right
)^2 \exp(-x) \right ], \nonumber
\end{eqnarray}
where $d_h$ is again the thickness of the hydrophilic layers
consisting of the water layer and the headgroups and
$\Delta_n=(\epsilon_{\mathrm{H}_2\mathrm{O}}(\omega_n)-\epsilon_{\mathrm{CH}_2}(\omega_n))/
(\epsilon_{\mathrm{H}_2\mathrm{O}}(\omega_n)+\epsilon_{\mathrm{CH}_2}(\omega_n))$
is a function of the frequency-dependent dielectric constants of
hydrocarbon and water. The prime symbol ${}^{\prime}$ indicates
that the static term $(n=0)$ has to be multiplied by $1/2$. The
calculation is somewhat involved, however Fenzl has shown that a
frequently used approximation of Eqs. \ref{vdW_petrache} is valid
for the dispersion term, but not for the static term which
dominates under salt-free conditions. Moreover, Podgornik and
coworkers have shown that nonpairwise additive contributions to
the van der Waals interaction play a significant role in
multilayers at large swelling \cite{Podgornik2006}. However, for
the present parameters, the above treatment should be sufficient.

Apart from the molecular forces discussed above, steric forces
resulting from membrane bending elasticity should be included, as
first introduced by Helfrich \cite{Helfrich}. Accordingly, a
repulsive undulation force arises
\begin{equation}
f_{U1} = 0.42 \frac{(k_B T)^2}{\kappa d_w^2} \label{fhelfrich1} ~,
\end{equation}
which cannot, however, be simply added to the molecular forces.
Instead, steric forces must be treated by field theoretical
approaches which go beyond the mean field approximation
\cite{Lipowsky}, or by self-consistent models \cite{Mecke}, but
which to date have not been combined with realistic molecular
potentials in multilamellar stacks. Facing these complications,
Petrache and coworkers \cite{Petrache} have pointed out that the
measured rms-fluctuation of the next neighbor distance $\sigma=
\sqrt{\eta} d/\pi$ can be used to experimentally determine the
fluctuation pressure $P_{fluct}$, which they then added to the
pressure calculated from the molecular potentials to fit their
data $\Pi(d) = P_{mol} +P_{fluct}$. Obviously, this approach
avoids the problematic identification of the thermodynamic
compression modulus to the bulk modulus $B$ as defined in the
Caill\'e model, but still assumes that the total pressure can be
written as a sum, which may strictly only be true in  mean field
approximation. The advantage of the approach is that it makes
clever use of the experimental information from either of the
interconnected functions $B(d)$, $\eta(d)$, or $\sigma(d)$ to
compute the pressure. According to \cite{Petrache}
\begin{equation}
P_{fluct}=-\frac{(4 k_B T)^2}{(8\pi)^2} \frac{1}{\kappa} ~\frac{d
\sigma^{-2}}{d~ (d_w)} ~. \label{pfluct}
\end{equation}
Fig. \ref{sigma} shows the measured parameter the inverse of the
fluctuation amplitudes $\sigma^{-2} (d)$, along with a fit to an
exponential decay (solid line). The data points can be fitted to
an exponential function $2.14 \, \exp (- d_w / 4.17) \un{\AA}^{-2}$.
Subtracting the corresponding fluctuation pressure $P_{fluct}$
obtained by differentiation according to Eq. \ref{pfluct} for a
given parameter $\kappa$ from $\Pi(d)$, the molecular interactions
(hydration and van der Waals interactions) can then be modelled
and compared to the data, as shown below in
Fig. \ref{petrache_fit}.

Below we give results for two different approaches in the data
analysis. The first approach is described in detail in
\cite{Ulidiss}. It is based on the assumption that the
periodicity $d$ is dominated by the molecular forces, and that
steric forces are comparatively small for relatively stiff
phospholipid membranes.

First approach: The calculation of the van der Waals interaction
was based on equation (\ref{VdWexakt}) for the static
contribution. The static part for $n=0$ was numerically integrated
between 0 and 100. For water
$\epsilon_{\mathrm{H}_2\mathrm{O}}(0)=80$ and for hydrocarbon
(tetradecane) $\epsilon_{\mathrm{CH}_2}(0)=2$ was taken. For the
dispersion term a hydrophobic bilayer thickness $26 \,
\mathrm{\AA}$ and a Hamaker constant $H_{\mathrm{dis}}=0.297$ was
used. The latter value has been chosen to approximate
(\ref{VdWexakt}), evaluated for dispersion relations
$\epsilon_{\mathrm{H}_2\mathrm{O}}(\omega)$ and
$\epsilon_{\mathrm{CH}_2}(\omega)$ which have been parameterized
by oscillator models as in \cite{Fenzl}. Note that in this
approach there is no free parameter for the vdW interaction.
Figure \ref{sketch} B shows the total interaction potential (solid
line) with separately calculated static and dispersion terms as
described above. A calculation based only on Eq. \ref{vdW_petrache}
(with adjusted Hamaker constant) is shown for comparison (dashed
curve). The parameters of the hydration interaction which were
then freely adjusted were \mbox{$H_0=4.8$ $k_BT/\mathrm{\AA}^2$}
and \mbox{$\lambda=1.88$ \AA}. The 10\% reduction from the fixed
values in \cite{Fenzl} in the van der Waals term for equation
(\ref{VdWexakt}) may be attributed to the fact that Helfrich
repulsive forces have not been included in the force balance. At
the same time, it is interesting to note the value obtained for
$\kappa$ in this approach. To estimate $\kappa$ we note that the
Caill\'{e} parameter $\eta = \frac{\pi}{2 d^2} \frac{k_BT}{\sqrt{B
\kappa /d}}$ has been determined from the full $q$-range fits to
the reflectivity curves at different pressures. As the compression
modulus $B(d)=\partial \Pi/\partial d$ has been independently
determined by numerical derivation of the measured $\Pi(d)$ curve,
one can now estimate the bending modulus $\kappa$ from the
experimental values of $\eta(d)$. The best agreement was obtained
for $\kappa = 23.2 \pm 2.5 \, k_BT$. Within these uncertainties
$\kappa \simeq 23 \, k_B T$ compares quite well with the value of
$\kappa = 19 \, k_B T$ at $30 \, ^{\circ}\mathrm{C}$ determined
from bulk suspensions of DMPC by Petrache {\em et al.}
\cite{Petrache}. Note however that another study employing full
$q$-range fits and osmotic pressure variation reports $\kappa =
11.5 \, k_B T$ \cite{Pabst03}, again at full hydration and
comparable $T$. Finally, thermal diffuse scattering analysis
points to significantly smaller values $\kappa = 7\, k_B T$
\cite{SaldittErratum}. Note that the determination of $\kappa$
from the osmotic pressure series is based on a problematic
assumption, i.e.\@ that the identification of the bulk modulus $B$
as defined in the Caill\'e model and the thermodynamic
compression modulus is correct. More details on the data analysis
following this approach based on molecular interactions only are
given in \cite{Ulidiss}.

\begin{figure}[htbp]
\includegraphics[width=9cm]{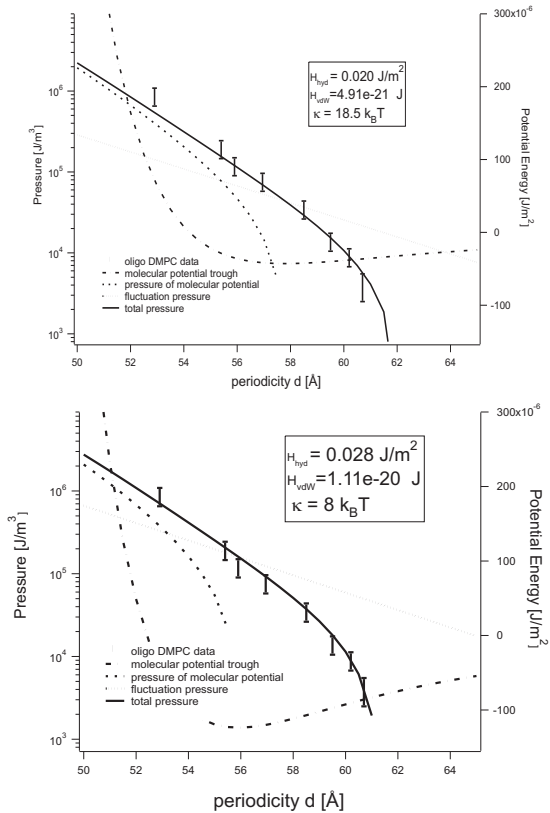}
\caption{Osmotic pressure $\Pi$ as a function of lamellar
periodicity $d$ (same data as in Fig. \ref{sigma}(A)). The simulations have
been carried out following the approach of \cite{Petrache} for (a)
fixed $\kappa=18.5 \, k_B T$ and (b) $\kappa=8 \, k_B T$. The parameters
of the hydration interaction $H_0$ (Eq. \ref{hydr}) and the Hamaker
constant $H_{vdW}$ (Eq. \ref{vdW_petrache}) are varied to match the
data. The simulations in (a) and (b)  show the total pressure
(solid line), the fluctuation pressure $P_{fluct}$ (dotted line),
the pressure corresponding to the molecular potential (dashed
line), and the potential trough corresponding to the molecular
forces (dash-dotted line). The fluctuation pressure as determined
from Eq. \ref{pfluct} and the data in Fig. \ref{sigma} was added to
the pressure stemming from the molecular forces.
\label{petrache_fit}}
\end{figure}

In the second approach we followed exactly the procedure given by
\cite{Petrache}. First, the fluctuation pressure is subtracted
from $P(d)$, and then the resulting bare pressure is modelled in
terms of the molecular interactions. However, the calculation of
the fluctuation pressure according to equation \ref{pfluct} needs
the bending rigidity $\kappa$ as an additional parameter. To
illustrate the range of parameter variability, we present a
comparison of two different choices of parameters: (a) all
parameters and functions are kept as close as possible to those
used in \cite{Petrache}, in particular keeping $\kappa=18.5 \, k_BT$
fixed. The corresponding values for $H_0=0.020 \, \un{J/m^2}$ and
$H_{vdW}=4.91 \cdot 10^{-21} \un{J}$ are practically identical to the
values in \cite{Petrache}, showing that the same approach can
explain both bulk and thin film data. The same treatment has then
been carried out for a different choice of $\kappa=8 \, k_BT$. Again
the simulations can be brought into agreement with the data, but
only for a different set of parameters $H_0=0.028 \un{J/m^2}$ and
$H_{vdW}=1.1 \cdot 10^{-20} \un{J}$. Thus values at the lower and upper
range of the $\kappa$ values reported in the literature both lead
to reasonable agreement, indicating that extra information from
other experiments is needed to unambiguously determine the
potentials. The potentials are shown for the two cases, and can be
compared also to the potential in Fig. 1(b), derived from the data
analysis under the assumption that the fluctuation repulsion is
negligible.

\section{\label{Conc}Discussion and Conclusions}

In conclusion, we have presented an osmotic pressure experiment on
thin solid-supported lipid multilayers (oligo-membranes). The
x-ray reflectivity has been measured and modelled over the full
$q_z$-range up to $0.7 \, \mathrm{\AA}^{-1}$. From this analysis
fluctuation and structural parameters can be obtained, similar to
the lineshape analysis of bulk suspensions
\cite{Petrache,PabstPRE}. Solid-supported oligo-membranes offer
some advantages both over thick multilamellar films and the bulk
counterpart. Long range thermal fluctuations are not as strong as
for bulk samples, and the scattering can be probed up to
higher momentum transfer. Owing to the smaller number of
bilayers, destructive interference in-between the Bragg peaks is
not quite as strong as in thick stacks of several hundred
bilayers. To achieve satisfactory fits, two important effects were
taken into account: (i) the static defects leading to a decreasing
coverage of the bilayers from the substrate to the top of the
film, and (ii) the thermal fluctuations of the bilayers subject to
the boundary condition of a flat substrate \cite{constantin}.
While (i) most likely reflects non-equilibrium aspects of sample
deposition and/or equilibrium wetting properties (not further
analyzed here), (ii) is exploited to deduce interaction parameters
in the framework of linear smectic theory. The curve $B (d)$
derived from the osmotic pressure series $\Pi (d)$ is subsequently
modelled based on different interaction potentials. However, this
modelling cannot be carried out without assumptions or additional
theoretic arguments.

In the data analysis, we have presented two entirely different
approaches to illustrate how the determination of interaction
forces depends on the specific assumptions, theoretical arguments,
or extra information taken from other experiments.  The first
approach builds upon the rather strong assumption that steric
forces are negligible and that the derivative of the equation of
state $\partial \Pi(d)/
\partial d$ can be identified with the modulus $B(d)$ which
controls the thermal fluctuations. It then yields the parameters
of the hydration force necessary to balance the van der Waals
attraction at each given osmotic pressure. This approach also
gives a value for the bending constant from simultaneous
inspection of $B(d)$ and $\eta(d)$. However, the resulting $\kappa
\simeq 23 \, k_BT$, is probably an overestimation. The rather
large value may point to the fact that $B$ is underestimated
by the contribution of only the bare potentials. Adding a
fluctuation pressure would tend to increase $B$ and thus decrease
$\kappa=K \, d$. Note that this determination of $\kappa$ is
conceptually very different from a more direct assessment of
$\kappa$, e.g.\@ from the measurement of diffuse scattering.

The second approach includes the steric Helfrich undulation
forces. This contribution is a subtle issue for the following
reasons: (a) it has been shown by Lipowsky and coworkers that the
Helfrich term cannot be simply added to the molecular forces. If
one nevertheless uses a mean field approach, (b) the functional
form to be used as well as the numerical prefactor are still under
debate \cite{Schilling01}. Therefore, we have followed an idea of
Petrache et al. \cite{Petrache}, who have carried out an osmotic
pressure study on DMPC, which is the bulk analogue of the present
work. Calculating the partition function within the linear smectic
elasticity model, they have derived derived an expression for the
fluctuation pressure as a function of a measurable quantity,
namely the derivative of the fluctuation amplitudes, see
Eq. \ref{sigma}. In a mean field treatment, they add this pressure
to the bare pressure calculated from the interaction potentials
and fit the sum to the measured curve $\Pi(d)$. In this step, an
assumption of $\kappa$ has to be made, e.g.\@ from other
experimental data. This approach has been carried out for two
choices of $\kappa$, see Fig. \ref{petrache_fit}.

We point out, however, that the questions related to the
interaction potentials arise only on a secondary level, where
structural results ($d$ and $\sigma$) are interpreted and
transformed to elasticity and interaction parameters. On the
primary level that the structural results presented here, {\em
i.e.} $\Pi(d)$, $\rho(z,\Pi)$ and $\left \langle u_n^2 \right
\rangle (d)$ are well supported by the curves and fits shown here.
The second level is necessarily model-dependent. We have presented
two alternative approaches to illustrate the relation and
interdependencies of different assumptions and results. It may be
justified to conclude that the second approach as proposed and
used by \cite{Petrache} is more appropriate, since steric
repulsion is known to be important. However, the choice of the
van der Waals expression may have to be improved to a more
accurate form,  and the choice of $\kappa$ is also an important
issue. Unfortunately, the second approach also relies on the
validity of a  mean field approximation. This simplification could
be eliminated in the future by generalization of a recent
self-consistent calculation for bilayer fluctuations and
interactions \cite{Mecke} to the case of several membranes or by
use of the approach developed in \cite{LipowskyNetz}. Furthermore,
non-linear effects due to the asymmetry of the potential could
also be included  by more general models
\cite{GaoGolubovic,LipowskyNetz} and/or numeric simulations. To
elucidate the validity of the linearized model {\em a posteriori},
the rms-deviation $\sqrt{\left \langle (u_n - u_{n+1})^2\right
\rangle}$ between neighboring membranes can be compared to the
width of the inter-bilayer potential well, see Figure \ref{sketch}
(B). For $N=16$ and $\eta=0.08$ (full hydration) the bilayers in
the center of the stack already exhibit considerable next-neighbor
distance fluctuations in the range of $4-5 \, \mathrm{\AA}$, when
compared to the water layer thickness. Thus the errors made in the
simplifying assumptions are probably not negligible. We note,
however, that at least under high osmotic pressure, where
fluctuations are small, both the mean field approach and the
harmonic approximation for the potential should hold.


\section*{Appendix}

In this appendix, we give a sketch of the derivation of our formula (\ref{R}) for the reflectivity,
insisting upon the separation between the specular and diffuse components. This is a classical result and a
more detailed derivation can be found in references \cite{sinha88} (equation 2.28) and \cite{AlsNiels}
(subsection 3.8.3) for the case of single interfaces and in \cite{sinha94} (section 3) for multiple
interfaces.

It is well known that bulk lamellar phases exhibit the Landau-Peierls instability, leading to a
characteristic power-law variation of the scattered signal \cite{Caille}. In such a system the fluctuation
amplitude $\langle u_n^2 \rangle$ diverges. It is then more appropriate to use the correlation of the
height difference, which remains finite for all finite values of $r$~:

\begin{eqnarray} \label{gmn}
g_{mn}({\mathbf r}_{\|})=\left \langle \left ( u_m({\mathbf r}_{\|})- u_n({\mathbf 0}) \right ) ^2 \right
\rangle\\
=\langle u_m^2 \rangle+\langle u_n^2 \rangle-2\langle u_m({\mathbf r}_{\|})u_n({\mathbf 0}) \rangle
\nonumber
\end{eqnarray}

It is then easy to show \cite{sinha88,sinha94} that the structure factor of the lamellar stack (without
taking into account the substrate contribution, so only the third term in Eq. (\ref{R}) is described)
reads~:
\begin{equation} \label{struct} S({\mathbf q}) = \sum_{m,n} \mathrm{e}^{-iq_zd(m-n)}
\int \dd {\mathbf r}_{\|} \, \mathrm{e}^{-i {\mathbf q}_{\|}{\mathbf r}_{\|}} \,
\mathrm{e}^{-\frac{1}{2}q_z^2g_{mn}(r)} \, .
\end{equation}
where $r=|{\mathbf r}_{\|}|$, assuming that the fluctuations are
isotropic in the membrane plane. For bulk systems, $\displaystyle
\lim_{r \rightarrow \infty} g_{mn}(r) =\infty$, so that
${\displaystyle \lim_{r \rightarrow \infty}} \exp \left [
-\frac{1}{2}q_z^2g_{mn}(r) \right ] =0$ and the Fourier transform
with respect to ${\mathbf r}_{\|}$ in formula (\ref{struct})
yields a "smooth" function $S({\mathbf q}_{\|})$ at fixed $q_z$.
If, however, $g_{mn}(r)$ does not diverge for $r \rightarrow
\infty$, the function
 $\exp \left [ -\frac{1}{2}q_z^2g_{mn}(r) \right ]$ now has a constant background, at a value of
 $\exp \left [ -\frac{1}{2}q_z^2g_{mn}(\infty) \right ] = \exp \left [ -\frac{1}{2}q_z^2
 \left ( \langle u_m^2 \rangle+\langle u_n^2 \rangle \right ) \right ]$, quantifying the "remanent order"
 in the system. Its Fourier transform is a Dirac delta function $\delta({\mathbf q}_{\|})$ (in practice, its width is
given by a combination of resolution effects, beam coherence and
system size). This term is sometimes called the "true specular
component", because the smooth function discussed above (the
"diffuse" component) also contributes to the specular signal
$S({\mathbf q}_{\|}={\mathbf 0},q_z)$. However, as the diffuse
scattering varies over a much larger ${\mathbf q}_{\|}$ range, it
can be accounted for in the first approximation by an offset scan
taken close enough to the specular sharp peak (see figure
\ref{fig:refl_pure_water}). Finally, we can write~:

\begin{equation}
S_{\mathrm{spec}}({\mathbf q}_{\|}={\mathbf 0},q_z) = \sum_{m,n} \mathrm{e}^{-iq_zd(m-n)} \, \mathrm{e}^{
-\frac{1}{2}q_z^2 \left ( \langle u_m^2 \rangle+\langle u_n^2 \rangle \right )}
\end{equation}
which is the form employed in equation (\ref{R}). It is noteworthy that this "true specular" contribution
is distinct in nature from the signal measured in SAXS experiments on powder samples, where only the
diffuse signal persists.

\begin{acknowledgement}

Guillaume Brotons is acknowledged for
helpful discussions on the osmotic stress technique. D. C. has
been supported by a Marie Curie Fellowship of the European
Community programme \textit{Improving the Human Research
Potential} under contract number HPMF-CT-2002-01903.

\end{acknowledgement}

\bibliography{OligoRef}

\begin{thebibliography}{10}

\bibitem{Lipowsky}
R.~Lipowsky.
\newblock Generic interactions of flexible membranes.
\newblock In R.~Lipowsky and E.~Sackmann, editors, {\em Handbook of Biological
  Physics}, volume~1, pages 521--602. Elsevier Science, Amsterdam, 1995.

\bibitem{Safinya}
C.R. Safinya, E.~B. Sirota, D.~Roux, and G.S. Smith.
\newblock {\em Phys Rev. Lett.}, 62:1134--1137, 1989.

\bibitem{Petrache}
H.I. Petrache, N.~Gouliaev, S.~Tristram-Nagle, R.~Zhang, R.~M. Suter, and J.F.
  Nagle.
\newblock {\em Phys. Rev. E}, 57:7014--7024, 1998.

\bibitem{petrache98}
H.~Petrache, S.~{Tristam-Nagle}, and J.F. Nagle.
\newblock {\em Chem. Phys. Lipids}, 95:83, 1998.

\bibitem{PabstPRE}
G.~Pabst, M.~Rappolt, H.~Amenitsch, and P.~Laggner.
\newblock {\em Phys. Rev. E}, 62:4000--4009, 2000.

\bibitem{Caille}
A.~Caill\'{e}.
\newblock {\em C. R. Acad. Sci. Paris, S\'{e}r. B}, 274:891, 1972.

\bibitem{Zhang}
R.~Zhang, R.~M. Suter, and J.~F. Nagle.
\newblock {\em Phys. Rev. E}, 50:5047--5060, 1994.

\bibitem{Smith}
G.S. Smith, E.B. Sirota, C.R. Safinya, and N.A. Clark.
\newblock {\em Phys. Rev. Lett.}, 60:813, 1988.

\bibitem{Lyatskaya}
Y.~Lyatskaya, Y.~Liu, S.~{Tristram-Nagle}, J.~Katsaras, and J.F. Nagle.
\newblock {\em Phys. Rev. E}, 63:011907, 2000.

\bibitem{VogelPRL}
M.~Vogel, C.~M\"{u}nster, W.~Fenzl, and T.~Salditt.
\newblock {\em Phys Rev. Lett.}, 84:390--393, 2000.

\bibitem{SaldittPRL}
T.~Salditt, M.~Vogel, and W.~Fenzl.
\newblock {\em Phys Rev. Lett.}, 90:178101, 2003.

\bibitem{Liu04}
Y.~Liu and J.F. Nagle.
\newblock {\em Phys. Rev. E}, 69:040901, 2004.

\bibitem{Fragneto}
G.~Fragneto, T.~Charitat, F.~Graner, K.~Mecke, L.~Perino-Gallice, and
  E.~Bellet-Amalric.
\newblock {\em Europhys. Lett.}, 53:100--106, 2001.

\bibitem{spin}
U.~Mennicke and T.~Salditt.
\newblock {\em Langmuir}, 18:8172--8177, 2002.

\bibitem{Li}
T.~Salditt, C.~Li, A.~Spaar, and U~Mennicke.
\newblock {\em Eur. Phys J. E.}, 7:105--116, 2002.

\bibitem{Parsegian}
V.A. Parsegian, R.P. Rand, N.L. Fuller, and D.C. Rau.
\newblock Osmotic stress for the direct measurement of intermolecular forces.
\newblock In L.~Packer, editor, {\em Methods in Enzymology}, volume 127.
  Academic Press, New York, 1986.

\bibitem{Fenzl}
W.~Fenzl.
\newblock {\em Z. Phys. B}, 97:333--336, 1995.

\bibitem{Brotons03}
G.~Brotons, T.~Salditt, M.~Dubois, and Th. Zemb.
\newblock {\em Langmuir}, 19:8235--8244, 2003.

\bibitem{StanleyStrey}
C.~Stanley and H.~Strey.
\newblock {\em Macromolecules}, 36:6888, 2003.

\bibitem{Ligoure97}
C.~Ligoure, G.~Bouglet, G.~Porte, and O.~Diat.
\newblock {\em J. Phys. II}, 7:473, 1997.

\bibitem{Ligoure99}
F.~Castro-Roman, G.~Porte, and C.~Ligoure.
\newblock {\em Phys. Rev. Lett.}, 82:109, 1999.

\bibitem{AlsNiels}
J.~Als-Nielsen and D.~McMorrow.
\newblock {\em Elements of Modern X-Ray Physics}.
\newblock Wiley, Chichester, 2001.

\bibitem{poniewierski}
A.~Poniewierski and R.~Ho{\l}yst.
\newblock {\em Phys. Rev. B}, 47:9840--9843, 1993.

\bibitem{constantin}
D.~Constantin, U.~Mennicke, C.~Li, and T.~Salditt.
\newblock {\em Eur. Phys. J. E}, 12:283--290, 2003.

\bibitem{dewetting}
L.~Perino-Gallice, G.~Fragneto, U.~Mennicke, T.~Salditt, and F.~Rieutord.
\newblock {\em Eur. Phys. J. E}, 8:275--282, 2002.

\bibitem{Pabst04}
G.~Pabst, H.~Amenitsch, P.~Kharakoz, P.~Laggner, and M.~Rappolt.
\newblock {\em Phys. Rev. E}, 70:021908, 2004.

\bibitem{Rand1}
R.P. Rand.
\newblock {\em Annu. Rev. Biophys. Bioeng.}, 10:227, 1981.

\bibitem{Podgornik2006}
R.~Podgornik, R.~H. French, and V.~A. Parsegian.
\newblock {\em J. Chem. Phys.}, 124:044709, 2006.

\bibitem{Helfrich}
W.~Helfrich.
\newblock {\em Z. Naturforsch.}, 28c:693, 1973.

\bibitem{Mecke}
K.R. Mecke, T.~Charitat, and F.~Graner.
\newblock {\em Langmuir}, 19:2080--2087, 2003.

\bibitem{Ulidiss}
U.~Mennicke.
\newblock {\em Structure and Fluctuations of Solid-Supported Phospholipid
  Membranes (in German)}.
\newblock PhD thesis, G\"{o}ttingen University, 2003.

\bibitem{Pabst03}
G.~Pabst, J.~Katsaras, V.~A. Raghunathan, and M.~Rappolt.
\newblock {\em Langmuir}, 19:1716--1722, 2003.

\bibitem{SaldittErratum}
T.~Salditt, M.~Vogel, and W.~Fenzl.
\newblock {\em Phys Rev. Lett.}, 93:169903, 2004.

\bibitem{Schilling01}
T.~Schilling, O.~Theissen, and G.~Gompper.
\newblock {\em Eur.Phys.J.E}, 4:103, 2001.

\bibitem{LipowskyNetz}
R.R. Netz and R.~Lipowsky.
\newblock {\em Phys. Rev. Lett.}, 71:3596--3599, 1993.

\bibitem{GaoGolubovic}
L.~Gao and L.~Golubovi\'{c}.
\newblock {\em Phys. Rev. E}, 67:021708, 2003.

\bibitem{sinha88}
S.~K. Sinha, E.~B. Sirota, S.~Garoff, and H.~B. Stanley.
\newblock {\em Phys. Rev. B}, 38:2297--2312, 1988.

\bibitem{sinha94}
S.~K. Sinha.
\newblock {\em J. Phys. (France) III}, 4:1543--1557, 1994.

\end{thebibliography}
\bibliographystyle{unsrt}
\end{document}